\documentclass[sigconf]{acmart}

\usepackage{subfigure} 
\usepackage[ruled]{algorithm2e}
\usepackage{algorithmic}
\usepackage{wasysym}
\usepackage{enumitem}
\usepackage{multirow}

\AtBeginDocument{%
  }

\setcopyright{acmlicensed}
\copyrightyear{2025}
\acmYear{2025}
\acmDOI{10.1145/3696410.3714565}

\acmConference[WWW '25] {Proceedings of the ACM Web Conference 2025}{April 28--May 2, 2025}{Sydney, NSW, Australia.}
\acmBooktitle{Proceedings of the ACM Web Conference 2025 (WWW '25), April 28--May 2, 2025, Sydney, NSW, Australia}
\acmISBN{979-8-4007-1274-6/25/04}

\begin{document}

\title{MixRec: Individual and Collective Mixing Empowers Data Augmentation for Recommender Systems}

\author{Yi Zhang}
\orcid{0000-0001-8196-0668}
\affiliation{%
  \institution{Anhui University}
  \city{Hefei}
  \state{Anhui}
  \country{China}}
\email{zhangyi.ahu@gmail.com}

\author{Yiwen Zhang}
\orcid{0000-0001-8709-1088}
\authornote{Yiwen Zhang is the corresponding author.}
\affiliation{%
  \institution{Anhui University}
  \city{Hefei}
  \state{Anhui}
  \country{China}
}
\email{zhangyiwen@ahu.edu.cn}

\renewcommand{\shortauthors}{Yi Zhang and Yiwen Zhang}

\begin{abstract}
The core of the general recommender systems lies in learning high-quality embedding representations of users and items to investigate their positional relations in the feature space. Unfortunately, data sparsity caused by difficult-to-access interaction data severely limits the effectiveness of recommender systems. Faced with such a dilemma, various types of self-supervised learning methods have been introduced into recommender systems in an attempt to alleviate the data sparsity through distribution modeling or data augmentation. However, most data augmentation relies on elaborate manual design, which is not only not universal, but the bloated and redundant augmentation process may significantly slow down model training progress. To tackle these limitations, we propose a novel Dual Mixing-based Recommendation Framework (\textsf{MixRec}) to empower data augmentation as we wish. Specifically, we propose individual mixing and collective mixing, respectively. The former aims to provide a new positive sample that is unique to the target (user or item) and to make the pair-wise recommendation loss benefit from it, while the latter aims to portray a new sample that contains group properties in a batch. The two mentioned mixing mechanisms allow for data augmentation with only one parameter that does not need to be set multiple times and can be done in linear time complexity. Besides, we propose the dual-mixing contrastive learning to maximize the utilization of these new-constructed samples to enhance the consistency between pairs of positive samples. Experimental results on four real-world datasets demonstrate the advantages of \textsf{MixRec} in terms of effectiveness, simplicity, efficiency, and scalability.
\end{abstract}




\maketitle

\section{INTRODUCTION}
As an integral part of modern Internet platforms, recommender systems have garnered significant attention \cite{ricci2011introduction, he2017neural}. At the core of recommender systems lies the wisdom of the crowd, leveraging group behavior to filter out irrelevant information \cite{rendle2009bpr, he2017neural}. With the rise of representation learning, the development of recommender systems has kept pace, with various neural networks being employed to model high-quality embeddings for users and items  \cite{li2024simcen}. As a data-driven scientific application, recommender systems now rely on interaction data more than ever. However, for various reasons, it is often challenging for individual platforms to obtain comprehensive user data, leaving recommender systems frequently dealing with data sparsity \cite{wu2021self, zhang2023revisiting}.

It is against this backdrop that many researchers have sought to introduce Self-supervised learning (SSL) \cite{liu2021self} into recommender systems to mitigate the data sparsity problem. This contains two paradigms, Generative and Contrastive. Specifically, generative models \cite{liang2018variational, zhang2023revisiting} based on probabilistic modeling can reconstruct historical interactions, while the more easily implemented contrastive learning \cite{chen2020simple} (CL) has made significant strides in combating data sparsity through data augmentation. Through data augmentation, contrastive learning maximizes mutual information from different views of the same sample, providing additional supervision signals for the recommendation task \cite{wang2021understanding}. Consequently, research is increasingly focusing on exploring different types of data augmentation \cite{yu2023self}. Early explorations concentrated on augmenting graph \cite{you2020graph}, where the graph is derived from user-item interactions \cite{wu2021self}. Researchers construct different views of the original input by randomly masking nodes or edges \cite{cai2023lightgcl}. Subsequently, a new line of thought has emerged, focusing on achieving augmentation in the feature space. Examples include introducing noise for representations \cite{yu2022graph, yang2023generative}, or finding semantic neighbors through clustering \cite{lin2022improving, yang2023generative}, attention \cite{ren2023disentangled, zhang2024exploring}, or hierarchical mechanisms \cite{he2023candidate}.

Despite the impressive progress of these novel methods, we argue that existing research still faces several limitations. Firstly, much of CL-based methods tends to rely on manually crafted augmentation. Examples include the need to precisely control the ratio of censored raw data \cite{wu2021self} or carefully add noise to raw data or representations to avoid over-corrupting the original semantic information \cite{yu2022graph}. More critically, to leverage more sophisticated augmentation strategies without overly disrupting recommendation, many recent methods introduce more hyper-parameters to balance multi-optimization \cite{yu2022graph, yang2023generative, zhang2024recdcl}. Given the wide variability of data structures, researchers are often required to test with multiple hyper-parameter sets for different application scenarios. This not only increases the cost of trial and error, but also raises concerns about improving subsequent research. Although some methods adopt an alternative approach by achieving data augmentation through repeated sampling \cite{huang2021mixgcf, zhang2023empowering}, this strategy significantly increases warm-up and introduces unquantifiable sampling bias \cite{zheng2021disentangling, chen2023bias}.

Secondly, the data augmentation strategies proposed in many existing works substantially increase the time costs associated with model training and inference. In early research, commonly used data augmentation methods are often accompanied by multiple encodings and propagations to obtain different perspectives of the same input \cite{wu2021self, yu2022graph}. And in more recent research, many studies introduce the paradigm of learnable (or adaptive) augmentation \cite{ren2023disentangled, zhang2024exploring}, as well as unique feature-lookup mechanisms \cite{lin2022improving, yang2023generative} to identify suitable new views in the semantic space. In practice, these strategies typically lead to an exponential increase in model training and inference time, as multiple iterations of the base recommendation model's encoder are needed to construct representations of various views. As noted in previous pioneering work \cite{wu2021self}, the time complexity of the model after introducing graph augmentation-based CL is 3.7 times greater than that of the base model.

The final and most significant limitation is that many CL-based methods do not utilize these new samples efficiently. Specifically, the broad approach aims to enhance the consistency of a pair of positive samples while maintaining uniformity among the negative samples \cite{chen2020simple}. While effective, this approach is insufficient for fully utilizing each sample, particularly as it neglects to explore the varying characteristics of individual samples. If we intend to introduce a larger number of samples, we may encounter some challenges. In other words, the approach lacks scalability. In a nutshell, although data augmentation is essential for data-sparse recommendation scenarios, existing augmentation strategies often struggle with the dilemma. This prevents us from achieving desired augmentations, thereby limiting the overall recommendation performance. Therefore, we further investigate whether the following objectives can be achieved simultaneously:
\begin{itemize}[leftmargin=*]
\item[$\bullet$] How can we more easily implement data augmentation with minimal hyper-parameter tuning?
\item[$\bullet$] How can we achieve data augmentation without significantly increasing model complexity?
\item[$\bullet$] How can new samples obtained through data augmentation be utilized more effectively to enhance recommendation?
\end{itemize}

To tackle the above limitations, we propose a novel  Dual Mixing-based Recommendation Framework (\textbf{\textsf{MixRec}}) for recommender system. \textsf{MixRec} does not rely on any external strategies to construct new views of the original sample, instead using the samples themselves. Specifically, inspired by the mixing mechanism \cite{zhang2018mixup}, we propose the \textbf{Individual Mixing} and \textbf{Collective Mixing}, respectively. Individual mixing aims to create new views that are more closely aligned with the original sample while also incorporating information from other samples. Collective mixing, on the other hand, emphasizes a more holistic perspective. Such a mixing process is a convex combination, which can be achieved with linear complexity and requires only simple parameter tuning to create new samples. In addition, we propose the \textbf{Dual Mixing Contrastive Learning}, which fully leverages all available new samples to enhance the supervision signals for the recommendation task. Since the inputs to both individual and collective mixing are just the user/item embeddings learned from specific encoder, \textsf{MixRec} can be seen as an easy-to-integrate framework that can be appended to most embedding-based recommendation methods. The major contributions of this paper are summarized as follows:
\begin{itemize}[leftmargin=*]
\item[$\bullet$] We propose a recommendation framework \textsf{MixRec}, which contains two lightweight data augmentation strategies, individual mixing and collective mixing for recommendation task.
\item[$\bullet$] We further propose dual-mixing contrastive learning, which maximizes the consistency of positive sample pairs from two perspectives while utilizing more negative samples to provide additional supervision signals for the recommendation task.
\item[$\bullet$] We conduct comparison experiments with twenty related baseline methods across four real datasets to thoroughly validate the effectiveness of the proposed \textsf{MixRec}.
\end{itemize}

\begin{figure*}[h]
  \centering
  \setlength{\abovecaptionskip}{0.1cm}
\setlength{\belowcaptionskip}{0.1cm} 
  \includegraphics[width=\linewidth]{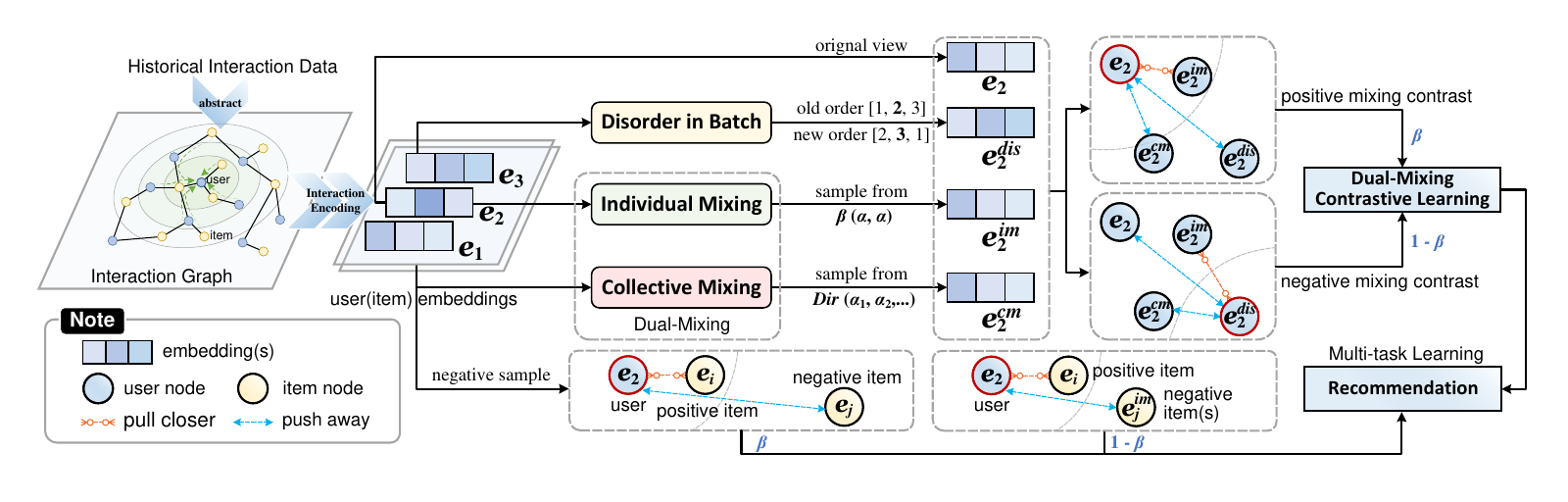}
  \caption{The complete information flow of the proposed \textsf{MixRec}. \textsf{MixRec} contains several phases of user-item interaction encoding, dual-mixing, dual-mixing contrastive learning, and multi-task learning for recommendation.}
  \label{fig_model}
\end{figure*}

\section{METHODOLOGY}
\subsection{Problem Formulation}
Without loss of generality, a recommendation scenario contains $M$ users ($\mathcal U=\{u_1, u_2, ..., u_M\}$) and $N$ items ($\mathcal U=\{i_1, i_2, ..., i_N\}$), and interactions between them \cite{he2017neural}. Based on existing work \cite{wang2019neural, he2020lightgcn}, historical interactions are typically stored as an interaction matrix $\mathbf R \in \mathbb R^{M\times N}$, where if there is an observed interaction between user $u$ and item $i$, we then have $R_{ui}=1$. The task of the recommender system aims to learn the prediction function as a means of predicting user $u$'s preference score $\hat{R}_{uj}$ for the item $j$ that have not been interacted with. For general recommendation scenarios, the only trainable parameters are the initial embedding representations for users $\{\mathbf e_u^{(0)}\}$ and items $\{\mathbf e_i^{(0)}\}$. Based on the above definition, we propose the dual mixing-based recommendation framework \textsf{MixRec}, as shown in Fig. \ref{fig_model}.

\subsection{User-Item Interaction  Encoding}
For ID-based recommender systems, interactions between users and items are crucial for modeling collaborative signals \cite{wang2019neural, he2020lightgcn}. A common approach is to use an encoder to obtain high-quality user/item embedding representations. Since the proposed \textsf{MixRec} is not restricted to any specific type of encoder, we adopt the state-of-the-art graph encoders \cite{kipf2016semi} to obtain user and item representations for consistency. Specifically, the user-item interaction graph \cite{wang2019neural} is denoted as $\mathcal G=<\mathcal U \cup \mathcal I, \mathcal E>$, where $\mathcal U$, $\mathcal I$, and $\mathcal E$ are the set of users, the set of items, and the interactions between users and items ($\mathcal E_{ui}=\mathcal E_{iu}=R_{ui}$). Given any user node $u \in \mathcal U$, the user embedding update process for an arbitrary layer is shown below:
\begin{equation}
\label{gcn}
\mathbf e_u^{(l)}=\text{agg}(\mathbf e_u^{(l-1)}, \{\mathbf e_i^{(l-1)}:i\in \mathcal N_u\}),
\end{equation}
where $\mathbf e_u^{(l)} \in \mathbb R^d$ is the embedding of user $u$ on the $l$-th layer, $d$ is the embedding size, $\mathcal N_u$ is the first-order neighbor set of user $u$, and $\text{agg}(\cdot)$ is a manually defined aggregation function. The item side has a similar definition. Considering that general recommendation only uses user and item ID as inputs, we adopt the widely used lightweight graph convolution \cite{he2020lightgcn} to encode user $u \in \mathcal U$ and item $i \in \mathcal I$ on the interaction graph:
\begin{equation}
\label{lightGCN}
\mathbf e_u^{\left( l \right)}=\sum_{i \in \mathcal N_u}p_{ui}\mathbf e_i^{\left( l-1 \right)}; \quad  \mathbf e_i^{\left( l \right)}=\sum_{u \in \mathcal N_i}p_{ui}\mathbf e_u^{\left( l-1 \right)},
\end{equation}
where $p_{ui}=1/\sqrt{|\mathcal N_u||\mathcal N_i|}$ is the graph Laplacian norm \cite{kipf2016semi, wang2019neural}. Compared to vanilla GCN \cite{kipf2016semi}, Eq. \ref{lightGCN} does not rely on feature transformation, and the process of information aggregation is done only by linear combination of neighbors, which is more in line with the semantic-free feature of implicit feedback \cite{rendle2009bpr,  he2017neural}. After $L$ layers of propagation, we construct usable embeddings for downstream tasks by addition \cite{zhang2024exploring}: $\mathbf e_{u}={\textstyle \sum_{l=1}^{L}\mathbf e_{u}^{(l)}}$ and $\mathbf e_{i}={\textstyle \sum_{l=1}^{L}\mathbf e_{i}^{(l)}}$.

\subsection{Dual-Mixing for Data Augmentation}

Although user and item representations obtained by interaction encoding can already be directly used in the downstream recommendation task, a serious problem is the data sparsity issuem\cite{zhang2023revisiting, wu2021self}. Therefore, data augmentation is especially important for recommendation task. Given that most existing data augmentation strategies for self-supervised contrastive learning \cite{chen2020simple, wu2021self} suffer from high complexity and a lack of flexibility, this dilemma prompts us to reconsider and rediscover a proven augmentation strategy. Our goal is to generate multiple new views as we wish, without relying on complex augmentation rules or time-consuming repetitive sampling. Building on this, we propose the \textbf{Dual-Mixing} for data augmentation, which incorporates individual and collective mixing.

\begin{figure*}
  \centering
    \setlength{\abovecaptionskip}{0.1cm}
\setlength{\belowcaptionskip}{0.1cm} 
  \includegraphics[width=\linewidth]{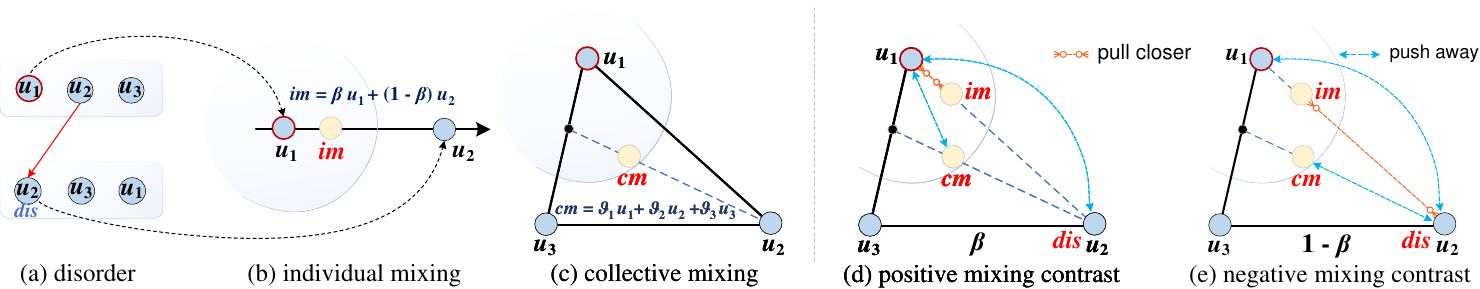}
  \caption{(a)-(c) Top examples of construction process for three new views, and the computation processes of (d) positive mixing contrastive loss $\mathcal L^{\text{pos}}_u$ and (e) negative mixing contrastive loss $\mathcal L^{\text{neg}}_u$ \textit{w.r.t.} user node $u_1$ (batch size $|\mathcal B|=3$).}
\label{fig_mixing}
\end{figure*}

\subsubsection{\textbf{Individual Mixing}} We first consider any individual (user $u$ or item $i$) in a sampled mini-batch. By interaction encoding, we have obtained their corresponding embedding representations: $\mathbf e_u$ and $\mathbf e_i$. Inspired by \cite{zhang2018mixup, kim2020mixco}, we construct new samples by linearly combining the original embedding pairs. Specifically, \textbf{individual mixing} creates synthetic training samples by linearly interpolating between two random samples from a sampled mini-batch:
\begin{equation}
\label{indi_mixup}
\mathbf e_u^{im}=\beta_u \cdot \mathbf e_u + (1-\beta_u) \cdot \mathbf e_u^{dis}; \quad
\mathbf e_i^{im}=\beta_i \cdot \mathbf e_i + (1-\beta_i) \cdot \mathbf e_i^{dis},
\end{equation}
where $\beta$ is a mixing coefficient drawn from a symmetric Beta distribution $Beta(\alpha, \alpha)$ with shape parameter $\alpha \in (0, \infty)$ (For simplicity, we define $\alpha=\alpha_u = \alpha_i$), and $\mathbf e_u^{dis}$ is the embedding at the corresponding position of user $u$ after the order of user embeddings in a batch has been disrupted (\textit{i.e.}, the embedding of any other user within the same batch).

Unlike traditional data augmentation, which is commonly used in contrastive learning \cite{you2020graph, yu2023self}, individual mixing does not require multiple instances of graph encoding or repeated sampling. More importantly, we can generate new views on demand by simply tweaking the shape parameter $\alpha$, while embedding $\mathbf e_u^{dis}$ is automatically obtained through a disordered operation. In practice, we empirically set $\alpha$ to 0.1, which eliminates the need for extensive manual effort in individual mixing. In addition, by controlling for a smaller $\alpha$, we constrain the sampled beta $\beta$ to yield larger values, enabling the newly generated sample $\mathbf e_u^{im}$ to retain as many properties of the original sample $\mathbf e_u$ as possible. This aligns with the invariance of contrastive learning \cite{li2022let}, treating these new sample as positive of the original view.

\subsubsection{\textbf{Collective Mixing}}
With individual mixing, we can easily generate new views for user $u$ or item $i$.  However, we note that the essence of recommender systems lies in the wisdom of the crowd \cite{he2017neural, he2020lightgcn}, highlighting the importance of considering inter-user relationships. Therefore, we further propose \textbf{collective mixing}, building upon individual mixing, to create new views that incorporate group information for each user. Specifically, collective mixing generates new examples by forming convex combinations of pairs of examples from the entire batch:
\begin{equation}
\label{coll_mix}
\mathbf e_u^{cm}=\vartheta_1\mathbf e_1+\vartheta_2\mathbf e_2+\cdots +\vartheta_{|\mathcal B|}\mathbf e_{|\mathcal B|},\quad \text{s.t.} \sum_{o=1}^{|\mathcal B|}\vartheta_o=1.0, 
\end{equation}
where  $\mathbf e_u^{cm}$ is the new view for user $u$ generated through collective mixing, with a similar definition for the item side. $\{\vartheta_1, \vartheta_2, ..., \vartheta_{|\mathcal B|}\}$ is a set of coefficients for convex combination, which is sampled from a multivariate Dirichlet distribution \cite{bai2024multimodality}:
\begin{equation}
\{\vartheta_1, \vartheta_2, ..., \vartheta_{|\mathcal B|}\} \sim Dirichlet(\alpha_1, \alpha_2, ..., \alpha_{|\mathcal B|}),
\end{equation}
where $\{\alpha_i: i \in \mathcal B\}$ is a set of positive real numbers to parameterize, and $\mathcal B$ is a sampled mini batch.

We now sort out the proposed dual-mixing, and the process of constructing new views is presented in Fig. \ref{fig_mixing}. For individual mixing (Fig. \ref{fig_mixing}(b)), the new sample $im$ is a linear interpolation of the original inputs $u_1$ and $u_2$. For collective mixing (Fig. \ref{fig_mixing}(c)), the new sample $cm$ produced by the convex combination lies within the convex hull of the original inputs $\{u_1, u_2, u_3\}$. Due to the influence of multiple sets of shape arameters, the $cm$ is affected by the influence from a wider range of users, which makes it extremely limited although it contains information from the original sample $u_1$. Therefore, the sample $im$ obtained from individual mixing we regard as a \textbf{positive sample} of the original view $u_1$, while the sample $cm$ produced by collective mixing is a \textbf{hard negative sample} of the original view $u_1$. And the sample $dis$ (\textit{i.e.,} $u_2$) obtained by disorder (Fig. \ref{fig_mixing}(a)) is naturally a \textbf{easy negative sample} of the original view $u_1$.

\subsection{Dual-Mixing Contrastive Learning}

Taking the user side as an example, through the proposed dual-mixing, we obtain two new views of user $u$ (\textit{i.e.}, $\mathbf e_u^{im}$ and $\mathbf e_u^{cm}$), along with one additional view $\mathbf e_u^{dis}$ of other users derived from disorder. It is now time to consider how to utilize these new views to provide additional supervision signals for the main recommendation task. A reasonable approach is to apply widely-used infoNCE \cite{chen2020simple, yu2023self} to maximize the mutual information between positive samples:
\begin{equation}
\label{infonce}
\mathcal L_{u}^{\text{cl}}=\frac{\text{exp}(s(\mathbf e'_u, \mathbf e''_u)/\tau)}{ {\textstyle \sum_{v\in \mathcal U}}\text{exp}(s(\mathbf e'_u, \mathbf e''_v)/\tau)},
\end{equation}
where $\mathbf e'_u$ and $\mathbf e''_u$ are additional views of user $u$ via data augmentation, $s(\cdot, \cdot)$ is the cosine similarity function, and $\tau$ is a temperature parameter. The contrastive loss essentially encourages the alignment of positive views while pushing the positive view $\mathbf e'_u$ away from the negative sample view $\mathbf e''_v$, thereby making the feature space more uniform \cite{wang2020understanding, wang2021understanding}. However, we argue that the supervision signals provided by vanilla contrastive learning are insufficient, as it fails to account for the uniformity across a wider range of sample pairs. In addition, the standard infoNCE loss overlooks the original view, leading to inconsistency between the auxiliary task and the main recommendation task. Therefore, we shift our focus and leverage the multiple views constructed earlier to propose the \textbf{dual-mixing contrastive learning}. Given user $u$, we first define the positive mixing contrastive loss:
\begin{equation}
\mathcal L_{u}^{\text{pos}}=-\text{log}\frac{\text{exp}(s(\mathbf e_u, \mathbf e_u^{im})/\tau)}{ {\textstyle \sum_{v\in \mathcal U}}[\text{exp}(s(\mathbf e_u, \mathbf e_v^{dis})/\tau)+\text{exp}(s(\mathbf e_u, \mathbf e^{cm}_v)/\tau)]}.
\end{equation}
In contrast to the original infoNCE, the above equation directly utilizes the original view as an anchor node to align the main task with the auxiliary task. In addition, we expand the number of negative sample pairs by contrasting the original view with multiple negative samples, further optimizing the uniformity of the entire feature space. To fully leverage the efficacy of these views, we next present the negative mixing contrastive loss as a counterpart:


\begin{equation}
\mathcal L_{u}^{\text{neg}}=-\text{log}\frac{\text{exp}(s(\mathbf e_u^{dis}, \mathbf e_u^{im})/\tau)}{ {\textstyle \sum_{v\in \mathcal U}}[\text{exp}(s(\mathbf e_u^{dis}, \mathbf e_v)/\tau)+\text{exp}(s(\mathbf e_u^{dis}, \mathbf e^{cm}_v)/\tau)]}. 
\end{equation}
In this step, we boldly employ the negative view $\mathbf e_u^{dis}$ at the corresponding position, obtained through order perturbation, as the anchor node while maintaining the roles of the other views. This design aims to effectively measure the correlation between these views, thereby providing additional supervision signals for the recommendation task. Note that the positive sample $\mathbf e_u^{im}$ is derived from mixing the two anchor nodes $\mathbf e_u$ and $\mathbf e_u^{dis}$. Consequently, both contrastive losses include a measure of the alignment between $\mathbf e_u^{im}$ with the two anchor nodes. The similarity of the positive sample $\mathbf e_u^{im}$ to the two anchor nodes is governed by the mixing coefficient $\beta_u$. Therefore, we similarly utilize $\beta_u$ to determine the weight of the two contrastive losses described above:
\begin{equation}
\label{mix_user}
\mathcal L_{\text{user}}= \sum_{u \in \mathcal U}\beta_u\cdot \mathcal L_{u}^{\text{pos}}+(1-\beta_u)\cdot\mathcal L_{u}^{\text{neg}},
\end{equation}
where $\beta_u$ is the mixing coefficient used for individual mixing in Eq. \ref{indi_mixup}. And fig. \ref{fig_mixing} (d)-(e) illustrates the dual-mixing contrastive process from the perspective of user $u_1$. For the item side, we have a similar definition, where $\beta_i$ is defined by Eq. \ref{indi_mixup}:
\begin{equation}
\label{mix_item}
\mathcal L_{\text{item}}= \sum_{i \in \mathcal I}\beta_i\cdot \mathcal L_{i}^{\text{pos}}+(1-\beta_i)\cdot\mathcal L_{i}^{\text{neg}}.
\end{equation}


\subsection{Multi-task Learning}
The optimization of \textsf{MixRec} involves two components: the recommendation task and the auxiliary task. For the main recommendation task, we take the widely applied BPR loss \cite{rendle2009bpr} to make more variability between positive and negative samples:

\begin{equation}
\mathcal L_{\text{BPR}}^{\text{pos}}=\sum_{<u,i>\in \mathcal O^+, <u,j>\in \mathcal O^-} -\text{ln}\sigma(\mathbf e_u^\top \mathbf e_i-\mathbf e_u^\top\mathbf e_j),
\end{equation}
where $\sigma$ is the sigmoid function. $i \in \mathcal I$ is an item that user $u \in \mathcal U$ has interacted with, and $j \in \mathcal I$ is any uninteracted one, and both of them are sampled from a uniform distribution \cite{huang2021mixgcf}. $\mathcal O^+$ and $\mathcal O^-$ are the observed and unobserved interaction sets, respectively.

As mentioned earlier, the original interaction data is highly sparse, making it infeasible to achieve satisfactory results by relying solely on the BPR loss described above. Therefore, we further utilize the previously mentioned individual mixing to construct additional negative samples for each user: $\mathbf e_j^{im} =\beta_i \cdot \mathbf e_j + (1-\beta_i) \cdot \mathbf e_j^{dis}$,  where item $j$ is the original negative sample for user $u$, and $\mathbf e_j^{dis}$ is constructed by disorder. Thus, within a mini-batch, we can construct multiple new negative samples for user $u$, resulting in the set $\mathcal O_u^{-}$. Subsequently, we compute the pairwise ranking loss for user $u$ based on these new negative samples:
\begin{equation}
\label{mix_bpr}
\mathcal L_{\text{BPR}}^{\text{neg}}=\sum_{<u,i>\in \mathcal O^+} -\text{ln}\sigma(\mathbf e_u^\top \mathbf e_i-\sum_{j \in \mathcal O_u^{-}} \mathbf e_u^\top\mathbf e_j^{im}).
\end{equation}
Compared to the original BPR loss, the equation above further considers the distance relationship between positive sample $i$ and multiple constructed negative samples $\mathcal O_u^{-}$, which encourages item $i$ to remain closer to user $u$ in the feature space. Similarly, we balance the two BPR losses through a linear combination:
\begin{equation}
\label{main}
\mathcal L_{\text{main}}=\beta_i\cdot \mathcal L_{\text{BPR}}^{\text{pos}} + (1-\beta_i)\cdot\mathcal L_{\text{BPR}}^{\text{neg}}.
\end{equation}



Finally, to integrate the recommendation task with the auxiliary task, we employ a multi-task joint training strategy for optimization. The complete optimization objective of \textsf{MixRec} is defined as follows:
\begin{equation}
\mathcal L_{\text{\textsf{MixRec}}}=\mathcal L_{\text{main}} + \lambda_1\cdot(\mathcal L_{\text{user}}+\mathcal L_{\text{item}}) + \lambda_2\cdot\left \| \Theta \right \|_2^2,
\end{equation}
where $\lambda_1$ and $\lambda_2$ are hyper-parameters to trade off the magnitude of losses, and $\Theta = {\mathbf E^{(0)}}$ is the set of trainable model parameter.

\subsection{Time Complexity}
In this section, we analyze the time complexity of \textsf{MixRec}. Specifically, we set the number of nodes and edges of the interaction graph $\mathcal G$ to be $|\mathcal V|$ and $|\mathcal E|$, respectively. $L$ is the number of GCN layers, $d$ is the embedding size, and $|\mathcal B|$ is the batch size. Next, we present the key components that contribute to the time complexity:
\begin{itemize}[left=0pt]
\item \textbf{Interaction Encoding:} The time complexity of this component is in line with mainstream methods since we adopt the design of the classical LightGCN \cite{he2020lightgcn}. Therefore, the time complexity of this component is $O(2|\mathcal E|Ld)$.
\item \textbf{Dual-Mixing:} Recalling Eqs. \ref{indi_mixup} and \ref{coll_mix}, this component does not significantly increase the time complexity, as the mixing operation involves only an addition of embeddings rather than matrix multiplication.
\item \textbf{Dual-Mixing Contrastive Learning:} In this component, we need to compute the contrastive loss on the user and item side, respectively. Therefore, the time complexity of this component is $O(4(|\mathcal B|d + 2|\mathcal B|^2d))$.
\item \textbf{Recommendation Losses:} We adopt the widely used BPR loss \cite{rendle2009bpr} to optimize the recommendation task. In addition, we additionally introduced mixed negative samples. Therefore, the time complexity of this component is $O(2|\mathcal B|d + |\mathcal B|^2d)$.
\end{itemize}
In practice, when \textsf{MixRec} adopts a graph encoder \cite{he2020lightgcn}, the time complexity of \textsf{MixRec} comes mainly from interaction graph encoding since the batch size $|\mathcal B|$ is much smaller than the interaction scale $|\mathcal E|$. As a result, the actual complexity of \textsf{MixRec} is slightly higher than that of LightGCN \cite{he2020lightgcn} due to the additional computation of losses; however, it remains significantly lower than other methods based on self-supervised contrastive learning or multiple sampling. This is primarily because \textsf{MixRec} requires only linear time complexity for data augmentation, avoiding the need to perform graph encoding multiple times. Additionally, \textsf{MixRec} does not require extra time for sampling multiple negative samples.

\begin{table}[t]
\small
\setlength{\abovecaptionskip}{0.1cm}
\setlength{\belowcaptionskip}{0.1cm} 
  \caption{ Statistics of the datasets.}
  \label{dataset}
  \begin{tabular}{l|c|c|c|c}
    \hline
    \textbf{Dataset}&\textbf{\#Users}&\textbf{\#Items}&\textbf{\#Interactions}&\textbf{Sparsity}\\
    \hline
    \hline
    \textbf{Douban-Book}&13,024&22,347&792,062&99.72\%\\
    \textbf{Yelp}&31,668&38,048&1,561,406&99.87\%\\
    \textbf{Tmall}&47,939&41,390&2,357,450&99.88\%\\
    \textbf{Amazon-Book}&52,643&91,599&2,984,108&99.94\%\\
    \hline
  \end{tabular}
\end{table}

\begin{table*}
        \centering
\setlength{\abovecaptionskip}{0.1cm}
\setlength{\belowcaptionskip}{0.1cm} 
  \caption{Overall performance comparisons on Yelp, Amazon-Book, Tmall, and Douban-Book datasets. The results of \textsf{MixRec} are \textbf{bolded}, whereas the best baseline is \underline{underlined}. * denotes that the improvement is significant with a $p$-value < 0.001 based on a two-tailed paired t-test. Part of the results are duplicated from original papers for consistency.}
  \label{performance1}
  \begin{tabular}{l|l|cc|cc|cc|cc}
    \hline
    \multirow{2}{*}{\textbf{Model Name}}&\multirow{2}{*}{\textbf{Year}}&\multicolumn{2}{c|}{\textbf{Yelp}}&
    \multicolumn{2}{c|}{\textbf{Amazon-Book}}&
    \multicolumn{2}{c|}{\textbf{Tmall}}&\multicolumn{2}{c}{\textbf{Douban-Book}}\\
	\cline{3-10}		~&~&Recall@20&NDCG@20&Recall@20&NDCG@20&Recall@20&NDCG@20&Recall@20&NDCG@20\\
    \hline
    \hline
    MF \cite{rendle2009bpr} & UAI'09  & 0.0539 & 0.0439 & 0.0308 & 0.0239 &0.0547 & 0.0400 &0.1292 &0.1147\\
    \hline
    Mult-VAE \cite{liang2018variational} & WWW'18& 0.0584 & 0.0450 & 0.0407 & 0.0315 & 0.0740 & 0.0552 & 0.1670 & 0.1604\\
    CVGA \cite{zhang2023revisiting} & TOIS'23    & 0.0694 & 0.0571 & 0.0492 & 0.0379 & 0.0854 & 0.0648 & 0.1736 & 0.1650\\
    DiffRec \cite{wang2023diffusion} & SIGIR'23 & 0.0665 & 0.0556& 0.0514 & \underline{0.0418} & 0.0792 & 0.0612 & 0.1619 & 0.1661\\
    \hline
    NGCF \cite{wang2019neural} & SIGIR'19 &0.0560&0.0456&0.0342&0.0261&0.0629&0.0465&0.1376&0.1215\\
    LightGCN \cite{he2020lightgcn} & SIGIR'20 & 0.0639 & 0.0525 & 0.0411 & 0.0315 & 0.0711 & 0.0530&0.1504&0.1404\\
    MixGCF \cite{huang2021mixgcf} & KDD'21 & 0.0713 & 0.0589 & 0.0485 & 0.0378 & 0.0813 & 0.0611 & 0.1731 &  0.1685\\
    IMP-GCN \cite{liu2021interest} & WWW'21 & 0.0653 & 0.0531 & 0.0460 & 0.0357& 0.0729 & 0.0539 & 0.1725 & 0.1604 \\
    CAGCN* \cite{wang2023collaboration} & WWW'23 & 0.0711 & 0.0590 & 0.0506 & 0.0400& 0.0783    & 0.0581 & 0.1704 & 0.1667\\
    \hline
    SGL-ED \cite{wu2021self} & SIGIR'21 & 0.0675 & 0.0555 & 0.0478 & 0.0379 & 0.0738 & 0.0556&0.1633&0.1585\\
    NCL \cite{lin2022improving} & WWW'22     & 0.0685 & 0.0577 & 0.0481 & 0.0373& 0.0750 & 0.0553&0.1647&0.1539\\
    DirectAU \cite{wang2022towards} & KDD'22 & 0.0703 & 0.0583 & 0.0506 & 0.0406 & 0.0752 & 0.0576&0.1660&0.1568\\
    SimGCL \cite{yu2022graph} & SIGIR'22  & 0.0721 & 0.0601& \underline{0.0515} & 0.0414& \underline{0.0884} & \underline{0.0674}&0.1728&0.1671\\
    GraphAU \cite{yang2023graph} & CIKM'23  & 0.0691 & 0.0574 & 0.0508 & 0.0403& 0.0840 & 0.0625&0.1699&0.1633\\
    CGCL \cite{he2023candidate} & SIGIR'23 & 0.0694 & 0.0561 & 0.0482 & 0.0375& 0.0861 & 0.0650&\underline{0.1741}&0.1667\\
    VGCL \cite{yang2023generative} & SIGIR'23   & 0.0715 & 0.0587& 0.0506 & 0.0401& 0.0880 & 0.0670&0.1733&\underline{0.1689}\\
    LightGCL \cite{yang2023generative} & ICLR'23   & 0.0692 & 0.0571& 0.0506 & 0.0397& 0.0833 & 0.0637 &0.1570&0.1455\\
    SCCF \cite{wu2024unifying} & KDD'24 & 0.0701 & 0.0580 & 0.0491 & 0.0399& 0.0772 & 0.0580& 0.1711 &0.1639\\
    RecDCL \cite{zhang2024recdcl} & WWW'24 & 0.0690 & 0.0560 & 0.0510 & 0.0405& 0.0853 & 0.0632 & 0.1664 & 0.1526\\
    BIGCF \cite{zhang2024exploring} & SIGIR'24 & \underline{0.0729} & \underline{0.0602} & 0.0500 & 0.0398& 0.0876 & 0.0664&\underline{0.1741}&0.1682\\
    \hline
    \textbf{\textsf{MixRec} (Ours)} &   & \textbf{0.0740*} & \textbf{0.0612*} & \textbf{0.0541*} & \textbf{0.0433*}& \textbf{0.0900*} & \textbf{0.0686*} & \textbf{0.1778*}& \textbf{0.1712*}\\
    \textit{p}-values  & & 5.21e-6 & 3.97e-5& 2.07e-7 & 9.14e-6& 4.75e-5 & 1.25e-5 & 9.82e-6 & 4.44e-5\\
\hline
  \end{tabular}
\end{table*}

\section{EXPERIMENTS}

In this section, we perform experiments on four real-world datasets
to validate our proposed \textsf{MixRec} compared with state-of-the-art
recommendation methods.

\subsection{Experimental Settings}
\subsubsection{\textbf{Datasets}} To validate the effectiveness of \textsf{MixRec}, we adopt four widely used recommendation datasets: Douban-Book \cite{yu2022graph, yang2023generative}, Yelp \cite{he2020lightgcn,yu2022graph}, Amazon-Book \cite{he2020lightgcn, yu2022graph}, and Tmall \cite{ren2023disentangled, zhang2024exploring}, which are varied in field, scale, and sparsity. Detailed statistics for four datasets are presented in Table \ref{dataset}. For fair comparison, preprocessing of all datasets remains consistent with previous studies \cite{he2020lightgcn, wu2021self}.

\subsubsection{\textbf{Baselines}}
To validate the effectiveness of \textsf{MixRec}, we choose the following state-of-the-art methods for comparison experiment:
\begin{itemize}[leftmargin=*]
\item[$\bullet$] \textbf{Factorization-based method}: MF \cite{rendle2009bpr}.
\item[$\bullet$] \textbf{Generative methods}: Mult-VAE \cite{liang2018variational}, CVGA \cite{zhang2023revisiting}, DiffRec \cite{wang2023diffusion}.
\item[$\bullet$] \textbf{GCN-based methods}: NGCF \cite{wang2019neural}, LightGCN \cite{he2020lightgcn}, IMP-GCN \cite{liu2021interest}, MixGCF \cite{huang2021mixgcf}, and CAGCN* \cite{wang2023collaboration}.
\item[$\bullet$] \textbf{SSL-based methods}: SGL-ED \cite{wu2021self}, NCL \cite{lin2022improving}, DirectAU \cite{wang2022towards}, SimGCL \cite{yu2022graph}, GraphAU \cite{yang2023graph}, CGCL \cite{he2023candidate}, VGCL \cite{yang2023generative}, LightGCL \cite{cai2023lightgcl}, SCCF \cite{wu2024unifying}, RecDCL \cite{zhang2024recdcl}, and BIGCF \cite{zhang2024exploring}.
\end{itemize}

\subsubsection{\textbf{hyperparameter Settings}}
\label{hyper}
We implement \textsf{MixRec} in PyTorch\footnote{ https://github.com/BlueGhostYi/ID-GRec}. For a fair comparison, the embedding size and batch size of all models are set to 64 (excluding Mult-VAE \cite{liang2018variational}, DiffRec \cite{wang2023diffusion}, and RecDCL \cite{zhang2024recdcl}) and 2048, respectively. For all graph-based methods, the number of network layers was set to 3 \cite{he2020lightgcn} (excluding IMP-GCN \cite{liu2021interest}). The default optimizer is Adam \cite{kingma2014adam}, and initialization is done via the Xavier method \cite{glorot2010understanding}. We follow the suggested settings in the authors’ original papers and use a grid search to choose the optimum hyper-parameters for all baselines. For \textsf{MixRec}, we empirically set the temperature coefficient $\tau$ to be 0.2. The weight of contrastive learning $\lambda_1$ is set in the range of \{0.01, 0.5, 0.1, 0.2, ..., 2.0\}, and the weight of $L_2$ regularization $\lambda_2$ is set in $1\text{e}^{-4}$ by default. The default setting for the shape parameter $\alpha$ is 0.1. To assess the performance of Top-N recommendation, we employ two commonly used evaluation metrics: Recall@N and NDCG@N (N=20 by default), which are computed by the all-ranking strategy \cite{wang2019neural, he2020lightgcn, yu2022graph}.

\subsection{Performance Comparisons}
\subsubsection{\textbf{Overall Comparisons}}
Table \ref{performance1} shows the performance of \textsf{MixRec} and all baseline methods on four datasets. \textsf{MixRec} achieves the best recommendation performance over all baselines on all datasets. Quantitatively, \textsf{MixRec} improves over the best baselines \textit{w.r.t.} Recall@20 by 1.78\%, 5.05\%, 1.81\%, and 2.13\% on Yelp, Amazon-Book, Tmall, and Douban-Book datasets, respectively. The experimental results demonstrate the effectiveness and generalization of \textsf{MixRec}. We attribute the performance improvement to the proposed individual and collective mixing, which effectively achieves data augmentation and alleviates the data sparsity problem faced by recommender systems.

MixGCF \cite{huang2021mixgcf}, another recommendation model utilizing the mixing mechanism, achieves better recommendation performance than MF and LightGCN, further demonstrating the superiority of the mixing mechanism. However, MixGCF merely samples multiple negative instances for calculating the BPR loss, which not only introduces sampling bias but also fails to provide additional supervision signals for the recommendation task. Compared to all contrastive learning-based methods, \textsf{MixRec} consistently outperforms them. This can be attributed to the fact that traditional contrastive learning methods do not fully leverage both positive and negative samples. In contrast, \textsf{MixRec} introduces dual-mixing contrastive learning, which effectively evaluates the role of various negative samples. Moreover, \textsf{MixRec} avoids the need for multiple graph encodings, ensuring its time complexity remains relatively low.

\begin{table}
\small
    \centering
    \caption{Efficiency comparison on Tmall and Amazon-Book datasets \textit{w.r.t.} time/epoch (T/E), number of epochs, and total runtime (measured in seconds (s), minutes (m), hours (h)). }
    \label{tab:runtime}
    \begin{tabular}{l|ccc|ccc}
    \specialrule{0.75pt}{0pt}{0pt}
    \multirow{2}{*}{} & \multicolumn{3}{c|}{\textbf{Tmall}} & \multicolumn{3}{c}{\textbf{Amazon-Book}}  \\
    \cline{2-7} 
    & T/E 
    & epochs   
    & runtime
    & T/E     
    & epochs
    & runtime \\
    \hline
    \hline
    LightGCN
    &49.1s	&286 & 3h50m  
    &54.7s &423 & 6h26m
    \\
    IMP-GCN
    &224.5s	&220 & 13h43m	
    &357.2s &260 & 25h48m
    \\
    MixGCF
    &180.3s	&114 & 5h43m	
    &202.5s &89 & 5h
    \\
    \hline
    SimGCL
    & 132.4s & 24 & 53m	  
    & 167.5s & 21 & 58m	  
    \\
    CGCL
    & 105.5s & 76 & 2h14m
    & 147.3s & 59 & 2h25m
    \\
    BIGCF	
    & 56.7s & 40 & 38m
    & 71.1s & 42 & 50m
    \\
    \hline
    \textbf{\textsf{MixRec}-1}
    & 32.4s & 34 & \textbf{18m} 	
    & 35.3s & 26 & \textbf{15m}
    \\
    \textbf{\textsf{MixRec}-3}
    & 56.1s & 26 & \textbf{24m} 	
    & 61.5s & 19 & \textbf{19m}
    \\
    \specialrule{0.75pt}{0pt}{0pt}
    \end{tabular}
\end{table}

\begin{figure}
\setlength{\abovecaptionskip}{0.0cm}
\setlength{\belowcaptionskip}{0.0cm} 
\centering
\subfigure[Tmall]{\includegraphics[width=1.62in]{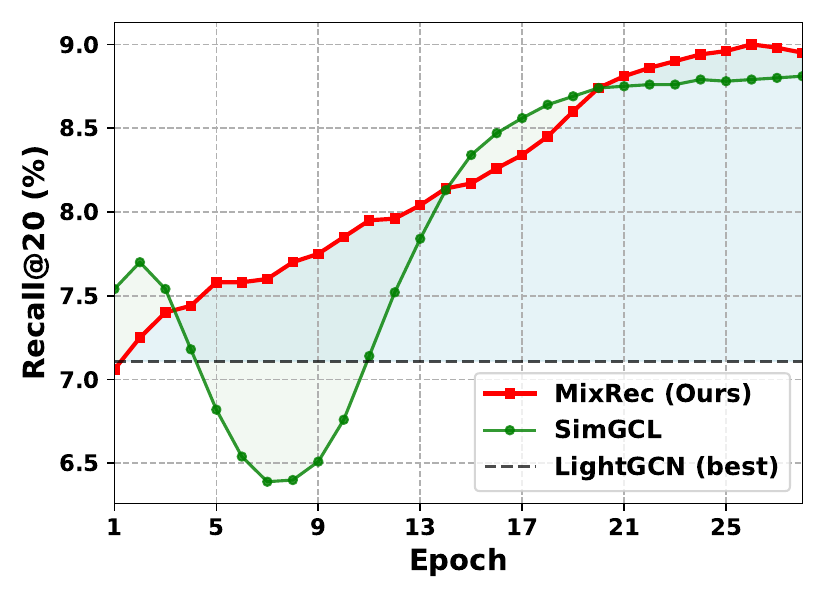}
\label{tmall_time_case}}
\hfil
\subfigure[Amazon-Book]{\includegraphics[width=1.62in]{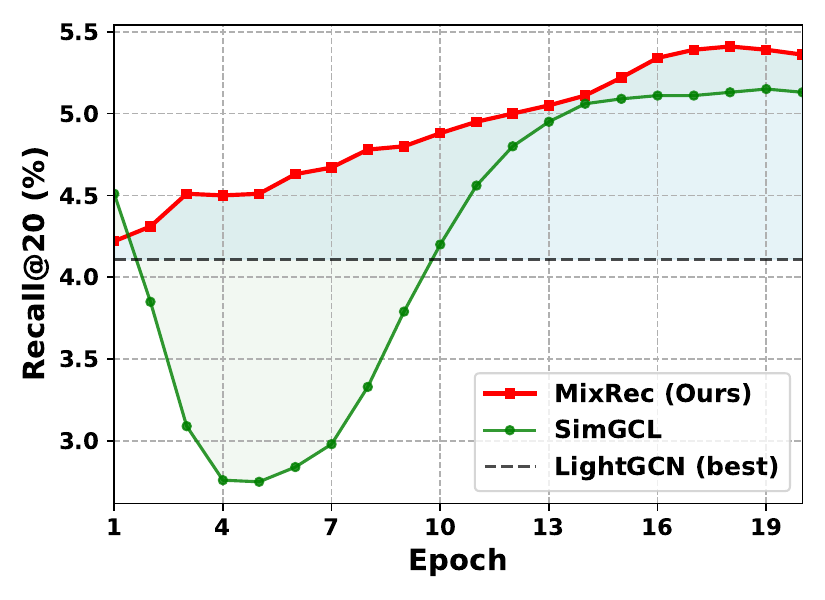}
\label{amazon_time_case}}

\caption{Training curves of LightGCN (best), SimGCL and \textsf{MixRec} on (a) Tmall and (b) Amazon-Book datasets.}
\label{fig_time}
\end{figure}

\subsubsection{\textbf{Comparisons w.r.t. Efficiency}} In this section, we present a comparison of the training time of \textsf{MixRec} ($L=1$ and $L=3$) with baseline methods on the two largest datasets, Tmall and Amazon-Book, as shown in Table \ref{tab:runtime}. As can be seen, while \textsf{MixRec}-3's single training time is only slightly higher than LightGCN \cite{he2020lightgcn} due to the additional loss computations, it is still significantly lower than other methods. \textsf{MixRec}-1, on the other hand, takes even less time than LightGCN. Besides, methods like LightGCN are constrained by sparse interaction data, requiring hundreds of iterations to achieve convergence. In contrast, \textsf{MixRec} converges in far fewer iterations, resulting in a significantly reduced overall training time. We further compare the training process of \textsf{MixRec} with the best-performing baseline method SimGCL \cite{yu2022graph} on Tmall and Amazon-Book datasets, as shown in Fig. \ref{fig_time}. Compared to SimGCL, \textsf{MixRec}'s training process is more stable, as evidenced by the absence of the performance drop seen in the early stages. We attribute this stability to \textsf{MixRec}'s ability to strike a better balance between alignment and uniformity, preventing early excessive uniformity from disrupting the effective distribution of the feature space \cite{wang2020understanding}.

\subsubsection{\textbf{Comparisons w.r.t. GCN Layers}} To maintain consistency in comparison, \textsf{MixRec} adopts a lightweight graph convolutional encoder \cite{he2020lightgcn}, consistent with previous works \cite{yu2022graph, zhang2024exploring}, to obtain user and item embedding representations. Table \ref{layer} provides comparisons of the effectiveness of \textsf{MixRec} and other methods with various GCN layer settings. \textsf{MixRec} maintains its performance advantage across all layers. And it should be noted that \textsf{MixRec} with only one layer outperforms even SimGCL \cite{yu2022graph} and BIGCF \cite{zhang2024exploring} with three layers, which shows \textsf{MixRec} can effectively mine user-item relationships without over-reliance on high-order graph structures, making it effective in reducing training costs (\textit{i.e.,} \textsf{MixRec}-1 in Table \ref{tab:runtime}).

\begin{table}[!t]

\setlength{\abovecaptionskip}{0.1cm}
\setlength{\belowcaptionskip}{0.1cm} 
\centering
\small
  \caption{Performance comparisons between \textsf{MixRec} and other baseline methods \textit{w.r.t.} number of GCN layer $L$.}
  \label{layer}
  \begin{tabular}{c|l|cc|cc}
    \hline
    \multirow{2}{*}{\textbf{\# Layers}}&\multirow{2}{*}{\textbf{Method}}&\multicolumn{2}{c|}{\textbf{Tmall}}&
    \multicolumn{2}{c}{\textbf{Amazon-Book}}\\
	\cline{3-6}	
    ~&~&R@20&N@20&R@20&N@20\\
    \hline
    \hline
    \multirow{3}{*}{$L=1$}&SimGCL \cite{yu2022graph} &0.0834&0.0635&0.0453&0.0358\\ 
    ~&BIGCF \cite{zhang2024exploring} &0.0851&0.0648&0.0466&0.0360\\ ~&\textbf{\textsf{MixRec}}&\textbf{0.0890}&\textbf{0.0681}&\textbf{0.0533}&\textbf{0.0429}\\
   \hline
       \multirow{3}{*}{$L=2$}&SimGCL \cite{yu2022graph}&0.0867&0.0665&0.0507&0.0405\\ 
    ~&BIGCF \cite{zhang2024exploring}&0.0865&0.0660&0.0493&0.0401\\
    ~&\textbf{\textsf{MixRec}}&\textbf{0.0896}&\textbf{0.0684}&\textbf{0.0535}&\textbf{0.0431}\\
   \hline
       \multirow{3}{*}{$L=3$}&SimGCL \cite{yu2022graph} &0.0884&0.0674&0.0515&0.0414\\ 
    ~&BIGCF \cite{zhang2024exploring}&0.0876&0.0664&0.0500&0.0398\\
    ~&\textbf{\textsf{MixRec}}&\textbf{0.0900}&\textbf{0.0686}&\textbf{0.0541}&\textbf{0.0433}\\
   \hline
\end{tabular}
\end{table}

\begin{figure}
\setlength{\abovecaptionskip}{0.0cm}
\setlength{\belowcaptionskip}{0.0cm} 
\centering
\subfigure[Tmall]{\includegraphics[width=1.62in]{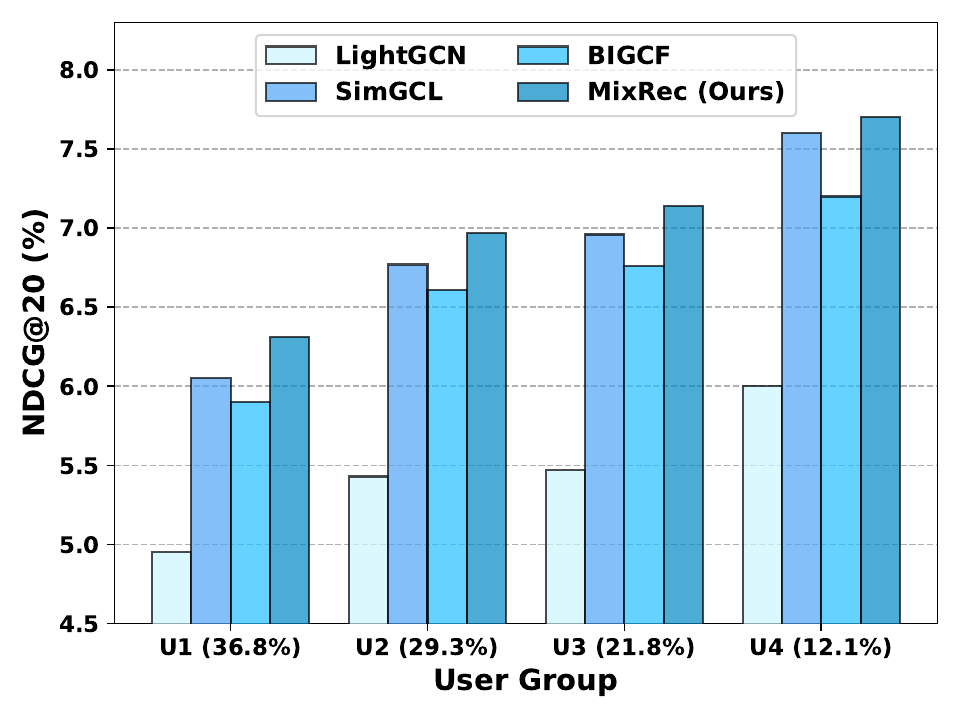}
\label{tmall_sparsity_case}}
\hfil
\subfigure[Amazon-Book]{\includegraphics[width=1.62in]{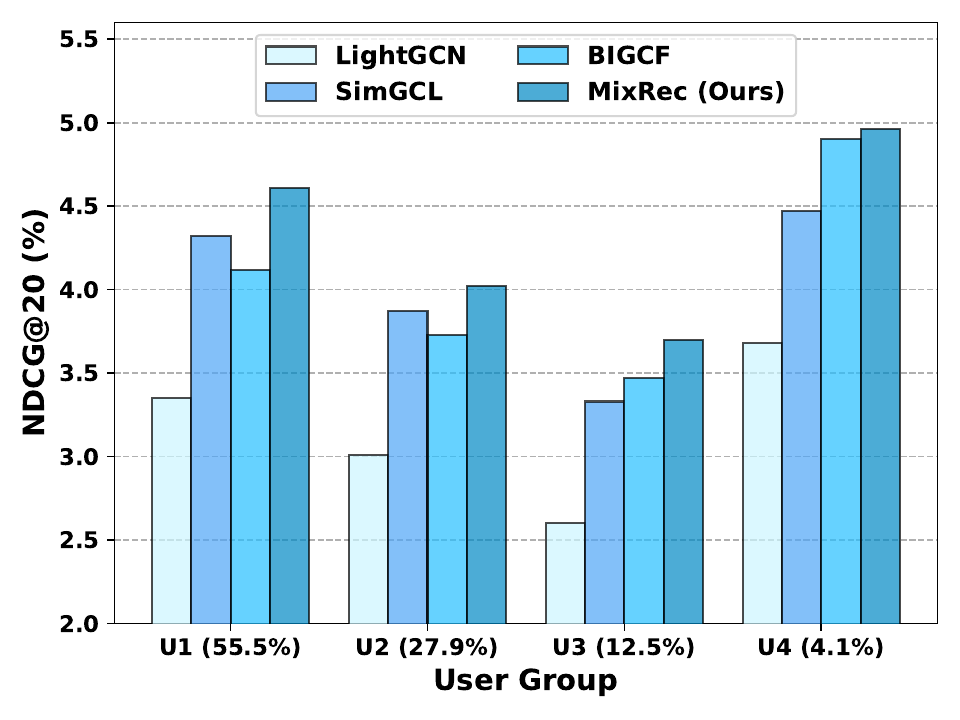}
\label{amazon_sparstiy_case}}

\caption{Sparsity tests on (a) Tmall and (b) Amazon-Book datasets. The $x$-axis shows user groups and proportions.}
\label{fig_sparsity}
\end{figure}

\subsubsection{\textbf{Comparisons w.r.t. Data Sparsity}}
In this section, we study the sparsity resistance of \textsf{MixRec} with the classical method LightGCN and the two well-performing baseline methods SimGCL and BIGCF on the two sparsest datasets, Tmall and Amazon-Book datasets. We take a generalized approach by dividing the interaction data from the training set evenly into four user groups $\{U_1, U_2, U_3, U_4\}$ based on the scale of the interactions. Specifically, $U1$ has the fewest interactions per user, indicating that it is the sparsest user group. And so on, $U4$ is the most active engaging user group. We train on the full training set and test each user group individually, and the experimental results are shown in Fig. \ref{fig_sparsity}. \textsf{MixRec} achieves noticeable performance gains across all sparse groups, further demonstrating its effectiveness. Focusing on the sparsest group $U1$, the improvement rates on the two datasets are 12\% and 10\%, respectively. We attribute this performance improvement primarily to the proposed individual and collective mixing strategies. These not only significantly increase the number of samples but also provide additional supervision signals for main recommendation task through dual-mixing contrastive learning.

\subsection{In-depth Studies of \textsf{MixRec}}
\subsubsection{\textbf{Ablation Studies}}
We construct a series of variants to verify the validity of each module in \textsf{MixRec}:
\begin{itemize}[leftmargin=*]
\item[$\bullet$] $\text{\textsf{MixRec}}_{\text{w/o DMCL (user)}}$: remove the Dual-Mixing Contrastive Learning on the user side (Eq. \ref{mix_user});
\item[$\bullet$] $\text{\textsf{MixRec}}_{\text{w/o DMCL (item)}}$: remove the Dual-Mixing Contrastive Learning on the item side (Eq. \ref{mix_item});
\item[$\bullet$] $\text{\textsf{MixRec}}_{\text{w/o IM}}$: remove individual mixing (Eq. \ref{indi_mixup}), and modify the positive sample to be the anchor node itself;
\item[$\bullet$] $\text{\textsf{MixRec}}_{\text{w/o DM}}$: remove collective mixing (Eq. \ref{coll_mix}).

\end{itemize}

\begin{table*}
        \centering

\setlength{\abovecaptionskip}{0.1cm}
\setlength{\belowcaptionskip}{0.1cm} 
  \caption{Ablation studies of \textsf{MixRec} on Yelp, Amazon-Book, Tmall, and Douban-Book datasets.}
  \label{tab:ablation}
  \begin{tabular}{l|cc|cc|cc|cc}
    \hline
    &\multicolumn{2}{c|}{\textbf{Yelp}}&
    \multicolumn{2}{c|}{\textbf{Amazon-Book}}&
    \multicolumn{2}{c|}{\textbf{Tmall}}&\multicolumn{2}{c}{\textbf{Douban-Book}}\\
	\cline{2-9}		&Recall@20&NDCG@20&Recall@20&NDCG@20&Recall@20&NDCG@20&Recall@20&NDCG@20\\
    \hline
    \hline
    w/o DMCL (user) (Eq. \ref{mix_user}) & 0.0676 & 0.0556 & 0.0484 & 0.0385 & 0.0826 & 0.0630 & 0.1566 & 0.1451\\
    w/o DMCL (item) (Eq. \ref{mix_item}) & 0.0652 & 0.0540 & 0.0436 & 0.0342 & 0.0815 & 0.0608 & 0.1593 & 0.1473\\
    w/o IM (Eq. \ref{indi_mixup}) & 0.0731 & 0.0605 & 0.0504 & 0.0403 & 0.0872 & 0.0665 & 0.1716 & 0.1628\\
    w/o DM (Eq. \ref{coll_mix}) & 0.0721 & 0.0597 & 0.0511 & 0.0409 & 0.0868 & 0.0664 & 0.1720 & 0.1639\\
    \hline
    \hline
    \textbf{\textsf{MixRec}} & \textbf{0.0740} & \textbf{0.0612} & \textbf{0.0541} & \textbf{0.0433}& \textbf{0.0900} & \textbf{0.0686} & \textbf{0.1778}& \textbf{0.1712}\\
\hline
  \end{tabular}
\end{table*}

The experimental results for all variants with \textsf{MixRec} on four datasets are shown in Table \ref{tab:ablation}. It is obvious that removing any of the modules resulted in varying degrees of performance degradation for \textsf{MixRec}, demonstrating the effectiveness of the various modules. Regarding the DMCL modules on both the user and item sides, the most significant performance degradation occurs when these modules are removed, indicating that relying solely on the main recommendation task is insufficient for modeling high-quality user and item embeddings. Focusing on the other three modules that utilize data augmentation, the observed performance decline further underscores the importance of data augmentation in mitigating the data sparsity problem.

\begin{figure}
\setlength{\abovecaptionskip}{0.0cm}
\setlength{\belowcaptionskip}{0.0cm} 
\centering
\subfigure[weight of loss $\lambda_1$]{\includegraphics[width=1.62in]{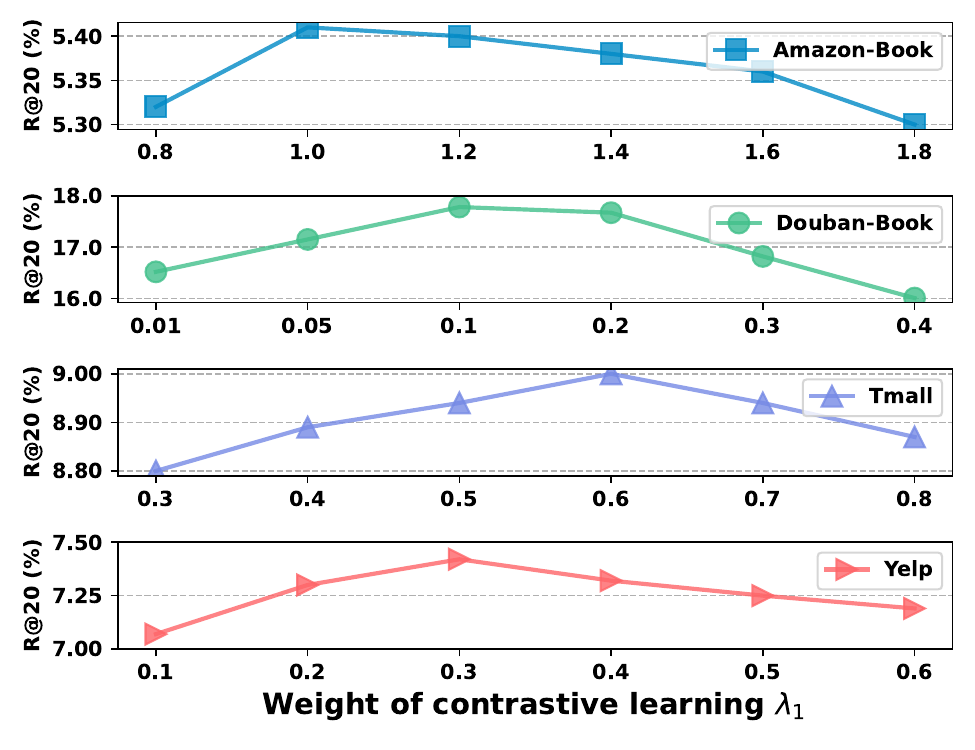}
\label{lambda_case}}
\hfil
\subfigure[shape parameter $\alpha$]{\includegraphics[width=1.62in]{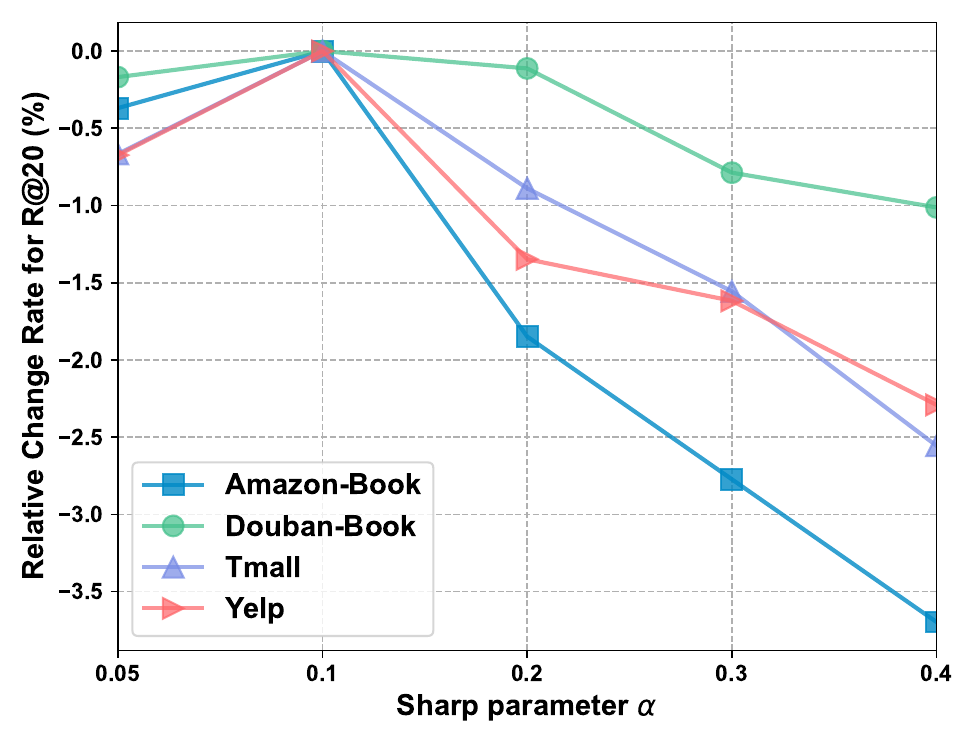}
\label{beta_case}}

\caption{Hyper-parameter sensitivities for (a) the weight of loss $\lambda_1$ and (b) shape parameter $\alpha$ on four datasets.}
\label{fig_param}
\end{figure} 

\subsubsection{\textbf{hyper-parameter Sensitivities}}
\label{hyper2}
Most of \textsf{MixRec}'s parameters are kept at the default settings (see Section 3.1.3 for details). Here we focus on two parameters $\lambda_1$ and $\alpha$. Their performance variations on four datasets are shown in Fig. \ref{fig_param}.
For the weight of contrastive losses, $\lambda_1$ shows different trends on four datasets. For dense datasets, $\lambda_1$ usually takes smaller values (\textit{e.g.}, Douban-Book); for sparse datasets, $\lambda_1$ takes larger values (\textit{e.g.}, Tmall and Amazon-Book). The optimal values on Amazon-Book, Douban-book, Tmall, and Yelp datasets are 1.0, 0.1, 0.6, and 0.3 respectively. 
For the shape parameter $\alpha$,  $\alpha = 0.1$ fetches the best recommendation performance in all cases. In other cases, too large a $\alpha$ will significantly degrade performance, indicating that the newly constructed sample is too corrupted to be considered a positive sample of the original view. Since $\alpha$ shows consistency across multiple datasets, we can set it to 0.1 by default without additional adjustment. Considering that all other hyper-parameters remain at the default values specified in Section \ref{hyper}, \textbf{the only hyper-parameter \textsf{MixRec} can adjust is the contrastive coefficient $\lambda_1$.}

\section{RELATED WORK}

\textbf{General Recommendation} General recommendation is a branch of recommender systems that focuses solely on user-item interactions \cite{ricci2011introduction}. In this context, implicit feedback is widely used due to its easy accessibility \cite{he2017neural}. However, improving the accuracy of greneral recommender system is challenging because of the sparsity of implicit feedback and the lack of semantic richness without auxiliary feature information \cite{zhang2023revisiting, li2024simcen}. Early work focuses on matrix factorization \cite{rendle2009bpr}, which eventually led to the introduction of neural networks to significantly enhance the model's learning capacity and generalization \cite{he2017neural}. With the rise of graph neural networks, researchers have begun abstracting historical interactions into bipartite graph to model high-order relationships between users and items \cite{ying2018graph, wang2019neural}. Among these novel efforts, LightGCN \cite{he2020lightgcn} has been widely adopted in subsequent research due to its ease of deployment and has replaced matrix factorization as a foundational model. The success of graph-based methods is evident not only in general recommendation scenario but also in other recommendation branches, including social network \cite{zhang2024dual}, knowledge graph \cite{wang2021learning}, and multimodal recommendation \cite{bai2024multimodality}, etc.

\noindent
\textbf{Self-supervised Learning for Recommendation}
Despite its considerable growth and a series of achievements, general recommendation still suffers from the sparsity of interaction data. Therefore, self-supervised learning is introduced into recommender systems to alleviate the data sparsity problem by constructing auxiliary tasks to provide additional supervision signals for the main recommendation task. In general, self-supervised learning can be broadly categorized into generative and contrastive models \cite{liu2021self}. The former aims to model the distribution of the user community and reconstruct the complete interaction at a probabilistic level, as seen in models like Mult-VAE \cite{liang2018variational} and DiffRec \cite{wang2023diffusion}. The latter leverages data augmentation to maximize mutual information between different views of the same sample. Earlier work focuses on modifying the input data \cite{cai2023lightgcl}, such as in SGL \cite{wu2021self}, which randomly masks edges or nodes on interaction graph. More recent work has shifted its focus toward finding new views within the feature space \cite{yu2023self}. Examples include introducing noise for representations \cite{yu2022graph, yang2023generative}, or finding semantic neighbors through clustering \cite{lin2022improving, yang2023generative}, attention \cite{ren2023disentangled, zhang2024exploring}, or hierarchical mechanisms \cite{he2023candidate}. Studies like DirectAU \cite{wang2022towards, yang2023graph, zhang2024recdcl} revisit CL from the perspectives of alignment and uniformity \cite{wang2020understanding}. SCCF \cite{wu2024unifying}, on the other hand, seeks to integrate graph convolution and contrastive learning into a unified framework. However, existing data augmentation strategies often suffer from high complexity and lack flexibility. More importantly, they typically measure the mutual information of a sample pair from only one perspective, failing to fully utilize the newly generated samples. \textsf{MixRec} is contrary to the design philosophy of pioneering works. Specifically, \textsf{MixRec} not only constructs richer new views with linear complexity but also maximizes the use of these samples through dual-mixing, enhancing its ability to support recommendation.

\section{CONCLUSION}
In this paper, we revisited the data sparsity problem entrenched in general recommendation and presented \textsf{MixRec}, an end-to-end recommendation framework based on dual-mixing, which contains individual mixing and collective mixing for data augmentation. Specifically, individual mixing aims to construct new samples that are unique to the original inputs with linear interpolation, while collective mixing considers the overall group perspective, creating new samples that represent the collective behavior of all users. Furthermore, we proposed dual-mixing contrastive learning to fully leverage all available sample pairs, maximizing the supervision signals provided for the recommendation task. We conducted extensive experiments on four real-world datasets and verified the effectiveness of \textsf{MixRec} in terms of recommendation performance, training efficiency, sparsity resistance, and usability.

\begin{acks}
This work is supported by the National Natural Science Foundation of China (No.62272001).
\end{acks}

\bibliographystyle{ACM-Reference-Format}
\balance
\bibliography{sample-base}

\newpage
\appendix

\section{APPENDIX}
In the Appendix, we first present the pseudo-code for the complete training of the proposed \textsf{MixRec}. Subsequently, Subsequently, we provide a detailed introduction to all comparative methods. Finally, we additionally provide experimental results to fully validate the effectiveness of \textsf{MixRec}.

\subsection{Algorithm Description}
The complete training procedure for the \textsf{MixRec} is described in Algorithm \ref{algorithms}. In the interaction encoding phase (line 5), \textsf{MixRec} obtains the embedding representations of users and items through an encoder. To construct multiple new views, \textsf{MixRec} needs to sample the corresponding mixing coefficients through the shape parameter $\alpha$ and randomly shuffle the arrangement of each batch $\mathcal B$ (lines 6-8). Subsequently, \textsf{MixRec} constructs new views through individual mixing (line 9) and collective mixing (line 10).
In the optimization stage (lines 11-12), we jointly optimize the dual-mixing contrastive loss and the main recommendation loss.

\begin{algorithm}
\caption{The training process of \textsf{MixRec}}
\label{algorithms}
\KwIn{user–item ineraction matrix $\mathbf R$, recommendation encoder $f_{\Theta}$; emperature coefficient $\tau$, weight of contrastive learning $\lambda_1$, regularization strength $\lambda_2$, and shape parameter $\alpha$.}
\begin{algorithmic}[1]
\STATE initialize parameters for $f_{\Theta}$;
\WHILE {\textsf{MixRec} not converge}
\STATE sample a mini-batch of user-item pairs $\mathcal B$;
\FOR{$<u,i,j>\in \mathcal B$}
\STATE construct embeddings $\mathbf e_u$, $\mathbf e_i$, and $\mathbf e_j$ by $f_{\Theta}$ (Eq . \ref{lightGCN});
\STATE sample $\beta_u$ and $\beta_i$ from $Beta(\alpha, \alpha)$;
\STATE sample $\{\vartheta_1, \vartheta_2, ..., \vartheta_{|\mathcal B|}\}$ from $ Dirichlet(\alpha_1, \alpha_2, ..., \alpha_{|\mathcal B|})$;
\STATE shuffle mini-batch $\mathcal B$ to obtain new view $\mathbf e_{u}^{dis}$ and $\mathbf e_{i}^{dis}$;
\STATE construct new views $\mathbf e_u^{im}$ and $\mathbf e_i^{im}$ by individual mixing (Eq. \ref{indi_mixup});
\STATE construct new views $\mathbf e_u^{cm}$ and $\mathbf e_i^{cm}$ by collective mixing (Eq. \ref{coll_mix});
\STATE calculate the dual-mixing contrastive losses $\mathcal L_{u}^{\text{pos}}$ and $\mathcal L_{u}^{\text{neg}}$ (Eqs. \ref{mix_user} and \ref{mix_item});
\STATE calculate the recommendation loss $\mathcal L_{\text{main}}$ by Eq. \ref{main};
\ENDFOR
\STATE average gradients from mini-batch;
\STATE update parameter by descending the gradients $\nabla_{\Theta}\mathcal L$;

\ENDWHILE
\RETURN model parameters $\Theta$;
\end{algorithmic}
\end{algorithm}

\subsection{Baseline Introduction}
\label{settings}
To verify the effectiveness of the proposed method, we compare \textsf{MixRec} with the following methods:
\begin{itemize}[leftmargin=*]
\item {\textbf{MF} \cite{rendle2009bpr}}: The method decomposes the interaction matrix into separate user and item embeddings and directly applies the inner product to compute the user's prediction score for each item.
\item {\textbf{Mult-VAE} \cite{liang2018variational}}: 
The method is a classical item-based CF model that captures both the mean and variance of user preferences. The model structure of Mult-VAE is set to [200, 600] as suggested in the source paper. The dropout rate range is in $\{0.1, 0.2, 0.3, 0.4, 0.5\}$ and the decay coefficient $\beta$ is in \{0.2, 0.4, 0.6, 0.8\}.
  \item {\textbf{CVGA} \cite{zhang2023revisiting}}: The method models user behavioral preferences as Gaussian distributions using GCNs and reconstructs the entire interaction graph through a multilayer perceptron. The decoder of CVGA uses a layer of neural network with a consistent dimension of the embedding size $d$, and the dropout rate is tuned in the range of $\{0.1, 0.2, 0.3, 0.4, 0.5\}$.
  \item {\textbf{DiffRec} \cite{wang2023diffusion}}: The method is an innovative item-based recommendation model that improves the model's robustness by progressively introducing Gaussian noise to the interaction data. The model structure is chosen from $\{[200, 600], [1000]\}$ with a step embedding size of 10. The diffusion and inference steps are $\{5, 40, 100\}$ and 0, respectively. The noise scale, minimum and maximum values are chosen from $\{1\text{e}^{-5}, 1\text{e}^{-4}, 1\text{e}^{-3}, 1\text{e}^{-2}\}$, $\{5\text{e}^{-4}, 1\text{e}^{-3}, 5\text{e}^{-3}\}$, and $\{5\text{e}^{-3}, 1\text{e}^{-2}\}$, respectively.
    \item {\textbf{NGCF} \cite{wang2019neural}}: The method uses a standard GCN design to encode embedding representations of users and items, while also taking into account user-item affinities. Additionally, we incorporate a message dropout mechanism for training.
  \item {\textbf{LightGCN} \cite{he2020lightgcn}}: The method is a classical GCN-based recommendation model that simplifies the original GCN design for CF tasks. The weight decay coefficient is set to $1\text{e}^{-4}$ by default, the number of GCN layers is set to 3.
  \item {\textbf{MixGCF} \cite{huang2021mixgcf}}: The method incorporates an additional hop mixing technique into LightGCN to generate negative samples, which are then used to compute the BPR loss. The setting range for the size of the candidate set for MixGCF is $\{16, 32, 64\}$.
  \item {\textbf{IMP-GCN} \cite{liu2021interest}}: The method is an enhanced GCN-based recommendation approach that partitions users into different subgraphs according to their interests. The number of GCN layers is set to 6, the number of user groups $g$ is set to 3 (2 on the Amazon-Book dataset due to the OOM issues), and the other settings are consistent with LightGCN \cite{he2020lightgcn}.
\item {\textbf{CAGCN*} \cite{wang2023collaboration}}: The method is an innovative GCN approach that constructs an asymmetric graph Laplacian matrix using a similarity matrix. The similarity calculation uses Salton Cosine Similarity, with the weighting coefficient ranging from \{1.0, 1.2, 1.5, 1.7, 2.0\}. and the other settings are consistent with LightGCN \cite{he2020lightgcn}.
\item {\textbf{SGL-ED} \cite{wu2021self}}: The method is a traditional GCL-based recommendation approach that generates different views of the same node by randomly dropping edges. The setting ranges for the contrastive loss coefficient, temperature, and edge mask probability of SGL-ED are \{0.005, 0.01, 0.05, 0.1, 0.5, 1.0\}, \{0.1, 0.2, 0.5, 1.0\}, and \{0.1, 0.2, 0.3, 0.4, 0.5\}, respectively.

\item {\textbf{NCL} \cite{lin2022improving}}: The method offers additional supervision signals for CF tasks by constructing both structural and semantic neighbors. The number of structural neighbor layers for NCL is set to 1, the weights for structural- and prototype-contrastive loss are set in the range of $\{1\text{e}^{-8}, 1\text{e}^{-7}, 1\text{e}^{-6}\}$, The weight of structure-contrastive loss on the item side are set in the range of \{0.1, 2.0\} (step size is 0.1), the temperature coefficients are set in the range of \{0.05, 1.0\} (step size is 0.025), and the number of clusters is set in the range of \{500, 800, 1000, 1500, 2000\} .
\item {\textbf{DirectAU} \cite{wang2022towards}}: The method reevaluates the collaborative filtering task through alignment and uniformity, aiming to optimize the distribution of users and items within the feature space. The range of adjustments for the weights controlling uniformity loss in DirectAU is \{0.2, 0.5, 1.0, 2.0, 5.0, 10.0\}.
\item {\textbf{SimGCL} \cite{yu2022graph}}: 
The method is a GCL-based recommendation approach that performs data augmentation by introducing random noise into the user and item embedding representations. The temperature coefficient of SimGCL is set to 0.2 by default. The weight of contrastive loss and noise strength are both set in the range of \{0.05, 0.1, 0.2, 0.5, 1.0\}.
        \item {\textbf{GraphAU} \cite{yang2023graph}}: The method tries to measure user-item alignment in each GCN layer and proposes a Layer-wise alignment pooling to measure the semantic distinctions among different layers. We manually adjust the weight vector $\alpha$ and the trade-off parameter $\gamma$ from 0.1 to 1.0.
        \item {\textbf{CGCL} \cite{he2023candidate}}: The method is a GCL-based recommendation approach that considers intermediate embeddings of different GCN layers to act as contrastive views. The three contrastive loss weights are searched in $\{1\text{e}^{-8}, 1\text{e}^{-7}, 1\text{e}^{-6}, 1\text{e}^{-5}, 1\text{e}^{-4}\}$, and we set the temperature coefficient $\tau=0.1$ by default. 
\item {\textbf{VGCL} \cite{yang2023generative}}: The method is an innovative GCL-based recommendation approach that learns user distributions through graph-based variational inference and constructs contrastive views via reparameterization trick and clustering. The weight of contrastive loss is set in the range of \{0.01, 0.05, 0.1, 0.2, 0.5, 1.0\}. The weight of cluster-level contrastive loss is set in the range of \{0.0, 1.0\} (step size is 0.1). The temperature coefficients are set in the range of \{0.10, 0.25\} (step size is 0.01). For the number of clusters of users and items, we use grid search to determine the optimal pair of values within the range of \{200, 400, 600, 800, 1000\}.
\item {\textbf{LightGCL} \cite{cai2023lightgcl}}: The method is based on GCL and constructs new views of the original interaction graph through singular value decomposition (SVD). The rank of SVD is set 5, and the weight of contrastive loss and temperature coefficient are both set from 0 to 1.
\item {\textbf{SCCF} \cite{wu2024unifying}}: The method attempts to combine graph convolution and the contrastive learning, using second-order cosine similarity for recommendation. The temperature coefficient is set in the range of \{0.1, 0.5\} (step size is 0.1).
\item {\textbf{RecDCL} \cite{zhang2024recdcl}}: The method attempts to combine batch-wise contrastive learning and feature-wise contrastive learning (FCL) for CF task. The weight coefficients of the two losses, $\alpha$ and $\beta$, range from \{0.2, 0.5, 1, 2, 5, 10\} and \{1, 5, 10, 20\}, respectively. The temperature coefficient is in the range of \{0.1, 0.3, 0.5, 0.7, 0.9\}. The weight coefficient for FCL loss ranges from \{0.005, 0.01, 0.05, 0.1\}.
\item {\textbf{BIGCF} \cite{zhang2024exploring}}: The method attempts to model user intentions based on the graph structure and regulates the feature space through graph contrastive regularization (GCR). The number of intentions is 128, and the temperature coefficient is set 0.2 by default. The weight of GCR loss is set in the range of \{0.0, 1.0\} (step size is 0.1).
\end{itemize}

\subsection{\textbf{Comparisons w.r.t. Base Models}} 
As we mentioned, since the inputs to both individual and collective mixing are just the original user/item embeddings, \textsf{MixRec} can be seen as an easy-to-integrate model that can be appended to most embedding-based recommendation methods. In the main comparison experiments, we adopt LightGCN \cite{he2020lightgcn} as the default encoder to ensure consistent comparison. To investigate the generalization ability of \textsf{MixRec}, we further select two types of base models: MF \cite{rendle2009bpr} and NGCF \cite{wang2019neural}. Specifically, the MF encoder does not incorporate graph information, while NGCF adopts the complete GCN design for CF task. We fine-tune the optimal hyper-parameters for each method, and the performance are shown in Table \ref{tab:base}. It can be observed that the performance improves significantly with \textsf{MixRec}. After introducing \textsf{MixRec}, the performance of MF, which is based solely on the inner product, has significantly improved, even surpassing graph-based methods such as NGCF and LightGCN. However, the benefit of MixRec for NGCF is moderate, possibly because the nonlinear transformation process of multiple GCN layers limits the further enhancement of NGCF's performance \cite{he2020lightgcn}.
\begin{table}
\centering
\small
\setlength{\abovecaptionskip}{0.1cm}
\setlength{\belowcaptionskip}{0.1cm} 
  \caption{Performance comparison of \textsf{MixRec} with different base models on Amazon-Book and Tmall datasets.}
  \label{tab:base}
  \begin{tabular}{l|ll|ll}
    \hline
    &
    \multicolumn{2}{c|}{\textbf{Amazon-Book}}&
    \multicolumn{2}{c}{\textbf{Tmall}}\\
	\cline{2-5} &Recall@20&NDCG@20&Recall@20&NDCG@20\\
    \hline
    MF \cite{rendle2009bpr} & 0.0308 & 0.0239 & 0.0547 & 0.0400\\
\textbf{w/ \textsf{MixRec}} & \textbf{0.0470}\tiny{+52.6\%} & \textbf{0.0371}\tiny{+55.2\%} & \textbf{0.0802}\tiny{+46.6\%} & \textbf{0.0610}\tiny{+52.5\%}\\
\hline
    NGCF \cite{wang2019neural} & 0.0342 & 0.0261 & 0.0629 & 0.0465 \\
\textbf{w \textsf{MixRec}} & \textbf{0.0415}\tiny{+21.3\%} & \textbf{0.0317}\tiny{+21.5\%} & \textbf{0.0705}\tiny{+12.1\%} & \textbf{0.0525}\tiny{+12.9\%}\\
\hline
    LightGCN \cite{he2020lightgcn} & 0.0411 & 0.0315 & 0.0711 & 0.0530\\
\textbf{w/ \textsf{MixRec}} & \textbf{0.0541}\tiny{+31.6\%} & \textbf{0.0433}\tiny{+37.5\%} & \textbf{0.0900}\tiny{+26.6\%} & \textbf{0.0686}\tiny{+29.4\%}\\
\hline
  \end{tabular}
\end{table}

\subsection{Optimal Hyperparameter Settings}
When using LightGCN \cite{he2020lightgcn} as the encoder, \textsf{MixRec} includes only two additional hyperparameters: the weight of the contrastive loss $\lambda_1$ and shape paramter $\alpha$. Compared to many recommendation methods that also incorporate self-supervised learning \cite{wu2021self, yu2022graph, zhang2024recdcl}, \textsf{MixRec} maintains a smaller number of hyperparameters. More importantly, in our previous study (Section \ref{hyper2}), we demonstrate that the shape parameter $\alpha$ exhibits good generalization (defaulting to 0.1). Therefore, \textsf{MixRec} only requires adjusting the weight of the contrastive loss $\lambda_1$. And the other hyperparameters can remain consistent with those of the chosen encoder. To enhance reproducibility, we list the hyperparameter settings of \textsf{MixRec} on four datasets, as shown in Table \ref{tab:best}.

\begin{table}
  \caption{Optimal hyperparameter settings of \textsf{MixRec} on Yelp, Amazon-Book, Tmall, and Douban-Book datasets. $|\mathcal B|$: batch size, $d$: embedding size, $\eta$: learning rate, $K$: the number of GCN layers, $\tau$: temperature coefficient, $\alpha$: shape parameter, $\lambda_1$: weight of contrastive loss, $\lambda_2$: weight of regularization.}
  \label{tab:best}
    \centering
    \begin{tabular}{l|c|c|c|c|c|c|c|c}
    \hline
        Dataset  & $|\mathcal B|$ & $d$ & $\eta$ & $K$ & 
 $\tau$ &$\alpha$ &$\lambda_1$ & $\lambda_2$\\ \hline
 \hline
        Yelp & 2048 & 64 & 0.001 & 3 & 0.2 & 0.1& 0.3 & 0.0001  \\
        \hline
        Amazon-Book & 2048 & 64 & 0.001 & 3 & 0.2 & 0.1& 1.1 & 0.0001  \\
        \hline
        Tmall & 2048 & 64 & 0.001 & 3 & 0.2 & 0.1& 0.6 & 0.0001  \\ 
        \hline
        Douban-Book & 2048 & 64 & 0.001 & 3 & 0.2 & 0.1& 0.1 & 0.0001  \\
        \hline
    \end{tabular}
\end{table}

\end{document}